\documentclass[showpacs,twocolumn,aps,amssymb,prl]{revtex4}
\usepackage{bm}
\usepackage{epsfig}

\begin{document}

\def\ba{{\bf a}}
\def\bk{{\bf k}}
\def\bp{{\bf p}}
\def\bq{{\bf q}}
\def\br{{\bf r}}
\def\bv{{\bf v}}
\def\bx{{\bf x}}
\def\bP{{\bf P}}
\def\bR{{\bf R}}
\def\bK{{\bf K}}
\def\bJ{{\bf J}}
\def\la{\langle}
\def\ra{\rangle}
\def\beq{\begin{equation}}
\def\eeq{\end{equation}}
\def\bea{\begin{eqnarray}}
\def\eea{\end{eqnarray}}
\def\bdm{\begin{displaymath}}
\def\edm{\end{displaymath}}
\def\nn{\nonumber}

\title{Dissipative dynamics of a harmonically confined 
 Bose-Einstein condensate}
\author{Z. Wu and E. Zaremba}
\affiliation{Department of Physics, Astronomy and Engineeering
Physics, Queen's University, Kingston, Ontario
 K7L 3N6, Canada.}
\date{\today}
\begin{abstract}
We study the dissipation of the centre of mass oscillation of a
harmonically confined condensate in the presence of a disorder
potential. An extension of the Harmonic Potential Theorem
allows one to formulate the dynamics from the point of view of an
oscillating disorder potential. This formulation leads to
a rigorous result for the damping rate in the limit of weak disorder.
\end{abstract}
\maketitle

Trapped Bose gases provide an ideal setting for the study of
nonequilibrium phenomena in a many-body system. Some examples
include condensate formation following a thermal
quench~\cite{miesner98,bijlsma00},
collective excitations as a function of
temperature~\cite{griffin09} and the
relaxation of highly nonequilibrium vortex
states~\cite{abo-shaeer02,jackson09}. In these, and
many other situations, the underlying superfluidity
plays an essential role in determining the dynamical
behaviour.

In some recent experiments~\cite{lye05,chen08,dries10}, the dissipative 
dynamics of a Bose
condensate in the presence of a disorder potential was studied.
This perturbing potential is the vehicle by which the collective
centre of mass motion of the condensate is dissipated by means
of internal excitations. The situation is analogous to the
motion of an impurity through a superfluid where it is found
that excitations can be produced above a critical
velocity~\cite{astrakharchik04}. Here
it is clear that the relative velocity of the impurity and
superfluid is the relevant variable; the motion of a heavy impurity
through a stationary superfluid or the flow of a superfluid past
a stationary obstacle are physically equivalent.

In this paper we demonstrate that a similar symmetry pertains
to a Bose gas trapped in a harmonic potential. Harmonic
confinement leads to an equivalence between the motion of the
condensate through a disorder potential that is at rest relative
to the confining potential, and the harmonic motion of the
disorder potential itself relative to the trapping potential and
the condensate. We exploit this equivalence to formulate a
rigorous theory of the centre of mass motion in the presence of
a disorder potential and obtain an estimate of the damping in
the limit of weak disorder. Our results are in qualitative
agreement with
those obtained earlier from an analysis of the 
one-dimensional Gross-Pitaevskii equation~\cite{albert}.

The Hamiltonian of the system studied experimentally is
\bea
&&\widehat H = \sum_{i=1}^N\left [ {\hat p_i^2 \over 2m} + V_{\rm
trap}(\hat \br_i)
\right ] + \widehat V_{\rm int} + \sum_{i=1}^N V_{\rm dis}(\hat \br_i)
\nonumber \\
&& \hskip .16truein \equiv \widehat H_0 + \widehat V_{\rm dis}
\label{Hamiltonian}
\eea
where $V_{\rm trap}(\br) = {1\over 2} m (\omega_\perp^2 \rho^2
+ \omega_z^2 z^2)$ is the trapping potential and $V_{\rm
dis}(\br)$ is
a disorder potential whose properties we specify later. The
interactions between the atoms is contained in $\widehat V_{\rm int}$.
To begin, we rephrase the Harmonic Potential Theorem
(HPT)~\cite{dobson94} in a form
which will be of particular utility in the subsequent development. 
Starting with the many-body state $|\Phi\rangle$, we
define the state
\beq
 |\Psi\ra=\exp\left \{\frac{i}{\hbar}\left
(\bp\cdot \widehat {\bR}-\bx\cdot\widehat {\bP}\right )\right
\}|\Phi\ra 
\equiv \widehat {T}(\bx,\bp)|\Phi\ra,
\eeq
where $\widehat {\bf R} = {1\over N} \sum_{i=1}^N \hat \br_i$
is the centre of mass coordinate and $\widehat {\bf P} = \sum_{i=1}^N
\hat \bp_i$ is the total momentum of the particles. These variables
satisfy the commutation relation $[\widehat R_\mu,\widehat P_\nu] =
i\hbar \delta_{\mu\nu}$. The unitary operator
$\widehat T$ shifts the state in position space by $\bx$ so that
$|\Psi(\br_1,...,\br_N)|^2 = 
|\Phi(\br_1-\bx,...,\br_N-\bx)|^2$. At the same time, the
state is shifted in momentum space by $\bp/N$ so that
$|\tilde \Psi(\bp_1,...,\bp_N)|^2 =
|\tilde \Phi(\bp_1-\bp/N,...,\bp_N-\bp/N)|^2$. This implies
that the total momentum is boosted by $\bp$. 

We now consider the time evolution of the state
$|\Psi\rangle$ according to the unperturbed Hamiltonian
$\widehat H_0$.
Using the Heisenberg equations of motion for the operators
$\widehat \bR$
and $\widehat \bP$ with respect to $\widehat H_0$, we find that
\beq
|\Psi(t)\rangle = e^{-i\widehat H_0 t/\hbar} |\Psi\rangle
= \widehat T(\bx(t),\bp(t)) e^{-i\widehat H_0 t/\hbar} |\Phi\rangle,
\eeq
%where
%\beq
%\widehat T(\bx,\bp,t) = e^{-i\widehat H_0 t/\hbar} \widehat
%T(\bx,\bp) e^{i\widehat H_0 t/\hbar}.
%\eeq
%Using the Heisenberg equations of motion for the operators
%$\widehat \bR$
%and $\widehat \bP$ with respect to $\widehat H_0$, we find that
%\beq
%\widehat T(\bx,\bp,t) = \widehat T(\bx(t),\bp(t)),
%\eeq
with
\bea
&&x_\mu(t) = x_\mu \cos\omega_\mu t + {p_\mu \over M\omega_\mu}
\sin\omega_\mu t \\
&&p_\mu(t) = p_\mu \cos\omega_\mu t - M\omega_\mu x_\mu
\sin\omega_\mu t, 
\eea
where $M= mN$ is the total mass of the system. If we now take
$|\Phi\rangle$ to be an eigenstate $|\Phi_\alpha\rangle$ of 
$\widehat H_0$, we see
that $|\Psi(\br_1,...,\br_N,t)|^2 =
|\Phi_\alpha(\br_1-\bx(t),...,\br_N-\bx(t))|^2$. In other words,
the probability density rigidly follows the motion of the centre
of mass of the system. This is essentially the content of
the HPT. More generally, the system
can be described by the density matrix
$\hat \rho = \sum_\alpha p_\alpha |\Phi_\alpha\rangle \langle
\Phi_\alpha|$.
If the states $|\Phi_\alpha\rangle$ are all eigenstates of
$\widehat H_0$, the total density of the system for the density
matrix $\widehat T(\bx,\bp)\hat \rho \widehat T^\dagger(\bx,\bp)$ 
oscillates rigidly
according to $n(\br,t) = n_0(\br-\bx(t))$ where
$n_0(\br)= {\rm Tr}(\hat \rho \hat n(\br))$. This applies to the 
special case of a thermal
equilibrium distribution. 
%We note in passing that the 
%ZNG theory of finite temperature dynamics satisfies the GHT~\cite{zaremba99}.

We next consider the dynamics of the centre of mass as governed
by the full Hamiltonian including the disorder potential. The
Heisenberg equations of motion lead to the equation
\beq 
{d^2Z \over dt^2} + \omega_z^2 Z = {F\over M},
\label{cofmeq}
\eeq
where $Z(t) = \langle \Psi(t)| \widehat R_z |\Psi(t) \rangle$
and
$F(t)=\langle \Psi(t)| \widehat F_z |\Psi(t) \rangle$ with
$\widehat F_z=
-\sum_{i=1}^N \partial V_{\rm dis}(\hat \br_i) /\partial \hat z_i
$.
Eq.~(\ref{cofmeq}) is an exact statement of the centre of mass
dynamics, but requires knowledge of the dynamical state $| \Psi(t) 
\rangle$. To
determine this state we go to the interaction picture and define
$|\Psi_I(t) \rangle \equiv \exp(i\widehat H_0 t/\hbar) |\Psi(t)
\rangle$ which satisfies
\beq
|\Psi_I(t) \rangle  = |\Psi(0) \rangle  - {i\over \hbar}
\int_0^t dt'\,\hat V_{{\rm dis},I}(t') |\Psi_I(t') \rangle
\label{eqofm_1}
\eeq
%\beq
%i\hbar {d|\Psi_I(t) \rangle \over dt} = \hat V_{{\rm dis},I}(t) |\Psi_I(t)
%\rangle,
%\eeq
with $\widehat V_{{\rm dis},I}(t) = \exp(i\widehat H_0 t/\hbar)
\widehat V_{\rm
dis} \exp(-i\widehat H_0 t/\hbar)$. 
%This equation has the formal solution
%\beq
%|\Psi_I(t) \rangle  = |\Psi(0) \rangle  - {i\over \hbar}
%\int_0^t dt'\,\hat V_{{\rm dis},I}(t') |\Psi_I(t') \rangle.
%\label{eqofm_1}
%\eeq

\begin{figure}[t]
\centering \scalebox{0.45}
 {\includegraphics{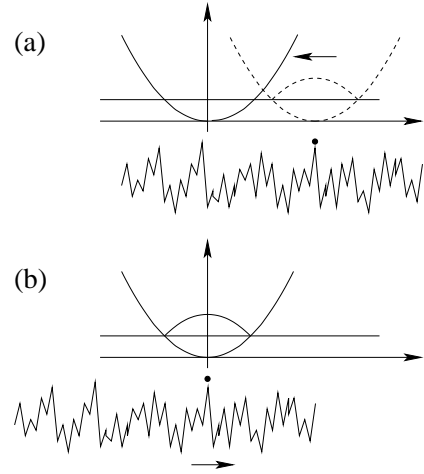}}
 \caption
 {(a) The condensate, originally in equilibrium with the
 unshifted trap (dashed) and the disorder potential, begins to
 oscillate about the centre of the shifted trap (solid).
 (b) The condensate, originally in equilibrium with the trap and 
 disorder potential, is driven by an oscillating disorder
 potential.
 }
\label{oscillation}
\end{figure}

In the experiment we consider~\cite{chen08}, 
%\textbf{the disorder potential is ramped on slowly to its full strength and then} 
the centre of mass
motion of the condensate is initiated by a sudden shift of the 
confining harmonic potential in the $z$-direction. To describe
this situation, we define the Hamiltonian of the system for
$t\le 0$ to be
\beq
\widehat H' =\sum_{i=1}^N\left [ {\hat p_i^2 \over 2m} + V_{\rm
trap}(\hat \br_i-{\bf x})
\right ] + \widehat V_{\rm int} + \sum_{i=1}^N V_{\rm dis}(\hat \br_i) 
%\nonumber \\
%&=&\widehat T (\bx,\bp)\widetilde H \widehat T^\dagger(\bx,\bp),
\label{Hprimed}
\eeq
while for $t > 0$, the system evolves according to the
Hamiltonian (\ref{Hamiltonian}). The trap potential in
(\ref{Hprimed}) 
%$\widehat H'$ 
is illustrated by the dashed curve in
Fig.~\ref{oscillation}(a); we assume that the state of the
system at $t=0$ is $|\Psi(0)\rangle = |\Psi_0\rangle$, the
ground state of $\widehat H'$. This Hamiltonian can be expressed
as $\widehat H' = \widehat T (\bx,\bp)\widetilde H \widehat
T^\dagger(\bx,\bp),$ where 
%the experimental initial conditions are 
$\bx=z_0\hat{\bf z}$, $\bp=0$ and $\widetilde H$ is
\beq
\widetilde H =\sum_{i=1}^N\left [ {\hat p_i^2 \over 2m} + V_{\rm
trap}(\hat \br_i)
\right ] + \widehat V_{\rm int} + \sum_{i=1}^N V_{\rm dis}(\hat \br_i+\bx).
\label{Htilde}
\eeq
The external potentials of this Hamiltonian are illustrated in
Fig.~\ref{oscillation}(b). The state
$|\widetilde \Psi_0\rangle=\widehat T^\dagger(\bx,\bp) |\Psi_0
\rangle$ is the ground state of $\widetilde H$.

Using the assumed initial state in (\ref{eqofm_1}), we find that
the state $|\widetilde \Psi_I(t) \rangle = \widehat T^\dagger(\bx,\bp) 
|\Psi_I(t) \rangle$ satisfies the equation
\beq
|\widetilde \Psi_I(t) \rangle = |\widetilde \Psi_0\rangle - {i\over \hbar}
\int_0^t \hskip -.05truein dt'\,
%\widehat T^\dagger (\bx,\bp) \widehat V_{{\rm dis},I}(t')
%\widehat T(\bx,\bp) 
\widetilde V_{{\rm dis},I}(\bx(t'),t')
| \widetilde \Psi_I(t') \rangle.
\label{eqofm_2}
\eeq
where
%The transformation of the disorder potential is
\bea
&&\hskip -.3truein \widetilde V_{{\rm dis},I}(\bx(t),t) \nn\\
&&\equiv
\widehat T^\dagger (\bx,\bp) \widehat V_{{\rm dis},I}(t)\widehat
T(\bx,\bp)\nn\\
%&&=\widehat T^\dagger (\bx,\bp) e^{i \widehat H_0 t/\hbar} \widehat V_{\rm
%dis}  e^{-i \widehat H_0 t/\hbar} \widehat T(\bx,\bp) \nn\\
&&=e^{i \widehat H_0 t/\hbar} \widehat T^\dagger (\bx(t),\bp(t))
\widehat
V_{\rm dis} \widehat T(\bx(t),\bp(t)) e^{-i \widehat H_0 t/\hbar}\nn\\
%&&\equiv \widetilde V_{{\rm dis},I}(\bx(t),t).
&& = e^{i \widehat H_0 t/\hbar}
\sum_{i=1}^N V_{\rm dis}(\hat \br_i + \bx(t)) e^{-i \widehat H_0 t/\hbar}.
\eea
%Since $\widehat V_{\rm dis}$ is a function of the particle
%coordinates, we find that
%\beq
%\widetilde V_{{\rm dis},I}(\bx(t),t) = e^{i \widehat H_0 t/\hbar}
%\sum_{i=1}^N V_{\rm dis}(\hat \br_i + \bx(t)) e^{-i \widehat H_0 t/\hbar}.
%\eeq
We thus see that $|\widetilde \Psi_I(t) \rangle$ is the
state that evolves from $|\widetilde \Psi_0\rangle$ as a result
of an {\it oscillating} disorder potential.

These results imply that the force appearing in (\ref{cofmeq})
can be expressed as
\beq
F(t) = \langle \Psi(t)|\widehat F_z | \Psi(t) \rangle = \langle
\widetilde
\Psi_I(t)| \widetilde F_{z,I}(t) | \widetilde \Psi_{I}(t)
\rangle,
\label{force}
\eeq
where
\beq
\widetilde F_{z,I}(t) = -e^{i \widehat H_0 t/\hbar}
\sum_{i=1}^N {\partial V_{\rm dis}(\hat \br_i + \bx(t))\over \partial \hat
z_i} e^{-i \widehat H_0 t/\hbar}.
\eeq
Eq.~(\ref{force}) is a key result and shows that there is an
intimate connection between the two very distinct physical
situations depicted in Fig.~\ref{oscillation}. 
In the first, one starts with an excited state
corresponding to a displaced condensate. This state then evolves
according to (\ref{eqofm_1}) in the presence of a static
disorder potential. Even though the condensate follows a damped
trajectory that eventually ends with the cloud being in
equilibrium with the static disorder, the total energy of the
system is conserved during this evolution. In the alternative
situation described by (\ref{eqofm_2}), the condensate starts in 
its ground state and is driven by a {\it dynamic} disorder
potential moving according to the unperturbed centre of mass
motion. In this case, the dynamic perturbation continually 
excites the condensate and the total energy increases as a
function of time. However, the atomic cloud eventually reaches a
steady state in which it moves with the disorder potential with no
further increase in energy. That the cloud experiences the
same force due to the disorder in these two situations is by no
means obvious and is a consequence of the fact that the system
is harmonically confined.

More quantitatively, the solution of (\ref{cofmeq}) for
%the state $|\Psi (t)\rangle$ with 
the assumed initial conditions is
%\beq
%Z(t) = z\cos \omega_zt + {p_z\over M\omega_z}\sin \omega_zt
%+ \widetilde Z(t),
%\label{Z(t)}
%\eeq
\beq
Z(t) = z_0\cos \omega_zt + 
%\widetilde Z(t) = 
{1\over M\omega_z} \int_0^t dt' \sin\omega_z(t-t')F(t'),
%\widetilde Z(t)
\label{Z(t)}
\eeq
where,
%\beq
%\widetilde Z(t) = {1\over M\omega_z} \int_0^t dt'
%\sin\omega_z(t-t')F(t').
%\eeq
because of (\ref{force}), the second term on the right hand 
side is in fact the displacement for the state
$|\widetilde \Psi_I (t)\rangle$. From (\ref{Z(t)}) we see 
that the change in the centre of mass position over one period $T =
2\pi/\omega_z$ is
\beq
\Delta Z_l \equiv Z(T_l) - Z(T_{l-1}) = -{1\over M\omega_z}
\int_{T_{l-1}}^{T_l} dt \sin\omega_zt F(t),
\label{deltaz}
\eeq
% In the experiments, the centre of mass
%motion is initiated by a sudden shift of the confining harmonic
%potential in the $z$-direction. At some point
%the disorder potential is then
%suddenly switched on. For concreteness, we choose the initial
%conditions $\bx = 0$ and $\bp = Mv_0\hat {\bf z}$, in which case
%(\ref{Z(t)}) gives for the centre of
%mass velocity $v(t) = dZ(t)/dt$ the result
%\beq
%v(t) = v_0 \cos\omega_zt+{1\over M}\int_0^t
%dt'\cos\omega_z(t-t') F(t').
%\eeq
%We see that the change in velocity over a period $T =
%2\pi/\omega_z$ is
%\beq
%\Delta v_l \equiv v(T_l) - v(T_{l-1}) = {1\over M}
%\int_{T_{l-1}}^{T_l} dt \cos\omega_zt F(t),
%\label{deltav}
%\eeq
where $T_l \equiv lT$. This is valid for both the static
and dynamic disorder potential scenarios.

To analyze the effects of $F(t)$ on the dynamics of the centre
of mass motion we will assume that the damping it gives
rise to is {\it weak}. In keeping with this assumption,
we evaluate $F(t)$ perturbatively. To second order in the 
disorder potential 
%with respect to the ground state of $\widehat H_0$ $|\Phi_0\rangle$, 
we have
%\bea
%|\widetilde \Psi_I(t) \rangle &\simeq& |\Phi_0\rangle - {i\over \hbar}
%\int_{-\infty}^0 dt'\, e^{\eta t'}\widetilde V_{{\rm dis},I}(\bx,t')| \Phi_0 \rangle \nonumber \\
%&&- {i\over \hbar}
%\int_0^t dt'\, \widetilde V_{{\rm dis},I}(\bx(t'),t')| \Phi_0 \rangle,
%\eea
%where $\eta$ is an positive infinitesimal number. This gives
\bea
&&\hskip -.3truein
F(t) = \langle \Phi_0 | \widetilde F_{z,I}(t) |
\Phi_0\rangle \nn\\&& \hskip .15truein
-{i\over \hbar} \int_{-\infty}^0 dt'\, e^{\eta t'}\langle \Phi_0 |[
\widetilde F_{z,I}(t),  \widetilde V_{{\rm dis},I}(\bx,t')] |
\Phi_0\rangle\nn\\&& \hskip .15truein
-{i\over \hbar} \int_0^t dt'\,\langle \Phi_0 |[
\widetilde F_{z,I}(t),  \widetilde V_{{\rm dis},I}(\bx(t'),t')] |
\Phi_0\rangle, 
\label{force_2}
\eea
where $|\Phi_0\rangle$ is the ground state of $\widehat H_0$.
The second term on the right hand side of (\ref{force_2})
involving the positive infinitesimal $\eta$ accounts for the
lowest order effect of the disorder on the ground state
$|\widetilde \Psi_0\rangle$, while the third term arises from
the dynamic perturbation of $|\widetilde \Psi_I(t)\rangle$ 
in (\ref{eqofm_2}).

We now write
\beq
\widetilde V_{{\rm dis},I}({\bf x}(t),t) = \int d\br \,V_{\rm dis}(\br+\bx(t)) \hat
n_I(\br,t),
\label{disorder}
\eeq
where $\hat n_I(\br,t)$ is the density operator in the
interaction picture. The disorder potential is represented as
\beq
V_{\rm dis}(\br) = \int {d\bk \over (2\pi)^3}
e^{i\bk\cdot\br} U(\bk),
\label{fourier}
\eeq
where the Fourier amplitudes $U(\bk)$ are stochastic variables 
having the following disorder averages:
\beq
\overline {U(\bk)} = 0,\quad
\overline {U(\bk)U^*(\bk')} = (2\pi)^3\delta(\bk-\bk')R(\bk).
\eeq
Inserting (\ref{disorder}) into (\ref{force_2}) and performing the
disorder average, we find that $\overline {F(t)} = \overline
{F_1(t)} + \overline {F_2(t)}$ with
\bea
&&\hskip -.15truein \overline {F_1(t)} = i \hskip -.05truein \int_{-\infty}^0 \hskip -.1truein dt' e^{\eta t'} 
\hskip -.075truein %\nonumber \\
\int \hskip -.05truein {d\bk \over (2\pi)^3}
R(\bk) k_z e^{i\bk\cdot[\bx(t) - \bx]} \chi(\bk,\bk;t-t'), \nonumber \\
%\label{avg_force}
%\eeq
%\beq
&&\hskip -.15truein \overline {F_2(t)} = i \hskip -.05truein 
\int_{0}^t dt'
\hskip -.05truein %\nonumber \\
\int \hskip -.05truein {d\bk \over (2\pi)^3}
R(\bk) k_z e^{i\bk\cdot[\bx(t) - \bx(t')]} \chi(\bk,\bk;t-t'), \nonumber \\
&&
\label{avg_force}
\eea
where $\chi(\bk,\bk;t-t')$ is the Fourier transform of the
density response function
\beq
\chi(\br,\br';t-t') = {i\over \hbar} \theta(t-t') \langle \Phi_0
| [\hat n_I(\br,t),\hat n_I(\br',t')]|\Phi_0
\rangle.
\label{drf}
\eeq
The disorder averaged force in (\ref{avg_force}) is the main
result of this paper and will be used to estimate the damping of
the centre of mass motion in the linear response regime.

In the experiments~\cite{chen08}, the speckle pattern is 
one-dimensional so that $R(\bk) = (2\pi)^2\delta(k_x)\delta(k_y) 
R(k_z)$ with
\beq
R(k_z) = \sqrt{\pi} \sigma
\overline{ V_{\rm dis}^2} e^{-\frac{1}{4}\sigma^2 k_z^2},
\eeq
where $\sigma$ is the correlation length of the gaussian
disorder and $\overline{ V_{\rm dis}^2}$ is the square of the standard
deviation of the disorder potential. In this situation, the 
response function of interest is
\beq
\chi(k_z,k_z;\tau) = \int d^3r \int d^3r'
e^{ik_z(z-z')}\chi(\br,\br';\tau).
\eeq
A formal expression for this quantity
can be given in terms of the exact Bogoliubov excitations of the
system. However, here we make use of a local density approximation
(LDA) whereby each element along the length of the condensate is
treated as part of a uniform cylindrical condensate having a
density per unit length of $\nu(z)$. In this approximation, 
the response function is taken to be 
\beq
\chi(k,k;\tau) \simeq \int dz \chi_{\rm cyl}(k,\tau;\nu(z)),
\label{LDA}
\eeq
where $\chi_{\rm cyl}(k,\tau;\nu(z))$ is the 
density response function of a uniform cylindrical condensate.
Here and in the following we drop the $z$ subscript on
$k_z$ for convenience. In the Bogoliubov approximation,
\beq
\hskip -.02truein \chi_{\rm cyl}(k,\tau;\nu(z)) = {i\over
\hbar}\theta(\tau)\sum_j \psi_j^2(k)\left
(e^{-i\omega_j(k)\tau} - e^{i\omega_j(k)\tau}\right ),
\label{chi_cyl}
\eeq
where $\psi_j(k) = 2\pi\int_0^\infty d\rho\,\rho \delta
n_j(\rho,k)$ is the cross-sectional average of the mode density 
fluctuation $\delta n_j(\rho,k)$; the index $j$
distinguishes the various radial modes of the cylindrical
condensate. 

We calculate the density fluctuation by treating the
condensate in the Thomas-Fermi (TF)
approximation~\cite{zaremba98}. This
is a good approximation in the experimental context since the
number of atoms in the cloud is of order $10^6$. 
In the hydrodynamic limit, the normalization of the
density fluctuation is then given by
$2\pi\int_0^{R_\perp} d\rho\,\rho \delta
n_j^2(\rho,k) = \hbar \omega_j(k)/2g$, where $R_\perp(z) =
\lambda \sqrt{R_z^2 -z^2}$ is the
transverse TF radius at the position $z$ along the axis. Here,
$\lambda = \omega_z/\omega_\perp$, $R_z =
\sqrt{2\mu/m\omega_z^2}$, $\mu$ is the chemical potential and
$g=4\pi a \hbar^2/m$.

It can be shown that the contribution of $\overline{F_1(t)}$ is
negligible in comparison to $\overline{F_2(t)}$. We thus focus on
the latter in the following. 
Substituting (\ref{LDA}) together with (\ref{chi_cyl}) into 
(\ref{avg_force}) we obtain
\bea
&&\hskip -.3truein \overline {F_2(t)} = -{1\over \hbar}
\int_{-R_z}^{R_z} dz \int \hskip -.05truein {dk \over 2\pi}
k R(k) \sum_j \psi_j^2(k)  \label{avg_force_2} \\
&& \hskip .175truein \times \int_0^t dt' e^{ik(z(t) - z(t'))} 
\left (e^{-i\omega_j(k)(t-t')} - {\rm c.c}\right ). \nn
\eea
For the initial conditions being considered, $z(t) =
z_0\cos\omega_zt$. In this case, we have
\beq
e^{ikz(t)} = \sum_{n=-\infty}^\infty e^{in\pi/2}
J_n(z_0k)e^{i\omega_nt},
\eeq
where $\omega_n = n\omega_z$ and $J_n(x)$ is the Bessel function
of the first kind of integral order $n$. This Fourier expansion
is substituted into (\ref{avg_force_2}) and the resulting
expression for $\overline {F_2(t)}$ is used in 
(\ref{deltaz}) to evaluate $\Delta Z_l$ explicitly.
%we obtain after a lengthy calculation
%\bea
%&&\hskip -.4truein \Delta v_l=-\frac {4\pi^2}{Mv_0(\hbar\omega_z)^2}
%\int_{-R_z}^{R_z} dz
%\int \frac {dk}{2\pi}R(k)\sum_j \hbar\omega_j(k)\psi_j^2(k) \nn \\
%&&\times \left \{ 
%\tilde{\Delta}_l(\nu)
%\bJ^2_{-\nu}(z_0 k)-\tilde{\delta}_l(\nu)\bJ_{-\nu}(z_0 k)
%\right \},
%\label {FT}
%\eea
%where $\nu = \omega_j(k)/\omega_z$. The various quantities
%appearing in (\ref{FT}) are the Anger function~\cite{watson52}
%\beq
%\bJ_\nu(x) = \frac{1}{\pi}\int_0^\pi d\theta \cos(x\sin\theta
%-\nu \theta),
%\eeq
%and
%\vskip -.25truein
%\begin{equation}
%\tilde{\Delta}_l(\nu)=\frac {\sin((2l-1)\pi \nu)}{\sin
%(\pi\nu)}, \quad
% \tilde{\delta}_l(\nu)=
%\frac{\sin((2l-1)\pi\nu)}{\pi\nu}
%\end{equation}
%In the $l\to \infty$ limit, $\tilde{\Delta}_l(\nu)$ behaves as
%$\sum_{n=-\infty}^\infty \delta(\nu-n)$ and
%$\tilde{\delta}_l(\nu)$ approaches $\delta(\nu)$.
Remarkably, we find that $\Delta Z_l$ is virtually independent
of $l$; the $l=1$ result differs from the $l\to \infty$ limit by
a few percent. There is essentially
no transient on the time scale of $T$ and implies that $Z(T_l)
\simeq z_0(1 -blT)$. 
Defining the damping of the oscillation as $bT = -\Delta
Z_\infty/z_0$,
we thus find 
\bea
&&\hskip -.5truein \frac{b}{\omega_z} = 
\frac{2\pi}{Mv_0^2\hbar} \int_{-R_z}^{R_z} dz
\int \frac {dk}{2\pi}R(k)\sum_j \psi_j^2(k) \nn \\
&& \hskip .5truein\times\sum_{n=1}^{\infty} n J^2_{n}(z_0 k)
\delta(\omega_j(k) - n\omega_z),
\label{damping_constant}
\eea
where $v_0 = \omega_z z_0$.

\begin{figure}[t]
\centering \scalebox{0.4}
{\includegraphics[50,220][560,600]{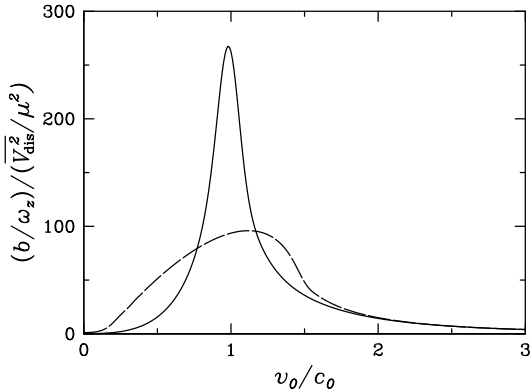}}
 \caption{The damping rate vs. $v_0/c_0$. The parameters used in
this calculation are~\cite{chen08}: $\omega_z/2\pi = 3.6$ Hz;
$\omega_\perp/2\pi = 180$ Hz; the $s$-wave scattering length $a=
200\,a_0$; $\mu/h = 1$ kHz; 
$\sigma = 10.6\,\mu{\rm m}$. The solid curve is for the
cylindrical LDA while the dashed curve is for the bulk LDA.
 }
\label{damping_rate}
\end{figure}
The results of our calculation of the damping rate
based on (\ref{damping_constant}) 
%as a function of $v_0/c_0$ 
are shown in Fig.~\ref{damping_rate}. We observe that the damping
rate exhibits a resonant peak at $v_0/c_0 \simeq 1$, where $c_0
= \sqrt{\mu/2m}$ is the sound speed in a cylindrical condensate
with chemical potential $\mu$~\cite{zaremba98}. 
Also shown in the figure
is the result obtained using the bulk LDA~\cite{brunello01} 
whereby each element
of the condensate is treated as a homogeneous gas. 
The cylindrical LDA is an improved approximation since it
explicitly accounts for the effect of the transverse confinement
on the excitations of the system and becomes exact in the limit
of a small aspect ratio $\lambda$ ($\omega_z \to 0$ with $\mu$
held fixed).

We now compare our results to the measurements presented in
Fig.~3 of Ref.~\cite{chen08}. The initial displacement of the
harmonic potential of $\sim$700 $\mu$m corresponds to 
$v_0/c_0 \simeq 2.9$. 
For the weakest disorder strength given of
$\overline {V_{\rm dis}^2}/\mu^2 = 0.0064$ we find
$(b/\omega_z)_{\rm th} \simeq 0.03$, whereas $(b/\omega_z)_{\rm exp}
\simeq 0.04$~\cite{chen08}. This should be taken as reasonably good
agreement given that there are no adjustable parameters in the 
calculation. In this regard, we emphasize that the damping rate
cannot be adequately characterized using a white-noise
spectrum~\cite{bhongale10}. 
%For a disorder strength of $\overline {V_{\rm dis}^2}/\mu^2 = 
%0.0256$ we find similar good agreement.
%it is crucial to take
%into account the experimental disorder correlation length
%$\sigma$~\cite{bhongale10}. %\textbf{The assumption of a white-noise spectrum can only be justified
%in the limit that the correlation length 
%is much smaller than than the healing length of the condensate.}~\cite{bhongale10}.

We have also analyzed the
data of Ref.~\cite{dries10} which is obtained using a different 
protocol to excite the centre of mass oscillation. Here, the
disorder is switched on \textit{suddenly} only after the oscillation of
the condensate has been initiated. 
%\textbf{as opposed to the experiments in Ref.~\cite{chen08} where the disorder potential is ramped on slowly to its full strength before the trap is shifted.} 
%This situation can also be
We find that the linear response
damping rate is still given by (\ref{damping_constant})
in this case. 
Using the experimental parameters corresponding to Fig. 7 of
Ref.~\cite{dries10}, we obtain $(b/\omega_z)_{\rm th} \simeq
0.06$ for $v_0/c_0 = 2.9$ and $\overline {V_{\rm dis}^2}/\mu^2
=0.0064$, whereas $(b/\omega_z)_{\rm
exp}\simeq0.002$~\cite{dries10}. 
We have no explanation for this discrepancy.
%We have not been able to identify
%an element of the theory that would account for this discrepancy.
Perhaps the analysis of a different situation such as
a gaussian perturbation [8] may shed light on
the limitations of the linear response calculation.

In summary, we have shown that the dissipative
dynamics of the centre of mass motion 
%of a harmonically confined condensate 
can be formulated in terms of a conventional response
function approach even though the initial state of the system is
far out of equilibrium. 
%This approach is quite general and,
%can be applied to arbitrary perturbations of the harmonic
%confinement as well as to finite temperatures. 
%With appropriate  modifications of the response function, this
With the appropriate response functions, this approach can also 
be used to study the dissipation at finite temperatures and in
fermionic systems.
%nonlinear corrections to the linear response results can also be
%addressed.

%We have no explanation for the
%discrepancy between theory and experiment but note
%that the theoretical value is consistent with the data in
%Ref.~\cite{chen08}
%when the smaller disorder correlation length of~\cite{dries10} 
%is taken into account.
%It would clearly be desirable to have additional experimental
%data in order to reconcile the differences between
%Refs.~\cite{chen08} and~\cite{dries10}, as well as to test the 
%limitations of the linear response caclulations.
%%and to verify the predicted dependence on the
%%various parameters such as $\mu$, $\sigma$, $a$ and $z_0$
%($v_0$).
%Possible extensions of the theory to
%finite temperature would also be of interest.
\vskip 8pt
\noindent{This work was supported by a grant from NSERC of
Canada.} We would like to acknowledge useful discussions with
Randy Hulet.

\end{document}